\documentclass[11pt,twoside]{book}
\usepackage{vshalo}
\usepackage{longtable}
\usepackage{amsmath,amssymb}
\usepackage{graphicx}
\usepackage{lscape}
\usepackage{psfig}
\usepackage{epsf}
\usepackage{index}
\usepackage{natbib}
\usepackage{bigdelim}
\usepackage{multirow}
\makeindex
\newindex{aut}{adx}{and}{\Large Author  Index}
\newcommand{\axindex}[1]{\index[aut]{#1}}

\begin{document}

\pagestyle{myheadings}
\setcounter{equation}{0}\setcounter{figure}{0}\setcounter{footnote}{0}
\setcounter{section}{0}\setcounter{table}{0}\setcounter{page}{35}
\markboth{Sterken, Samus \& Szabados}{VS-Halo Papers}
\title{Variable stars, distance scale, globular clusters}
\author{Alexey S. Rastorguev$^{1,2}$}
\axindex{Sterken, C.}\axindex{Samus, N. N.}\axindex{Szabados, L.}
\affil{$^1$Moscow State University, Faculty of Physics, Moscow, Russia\\
$^2$Sternberg Astronomical Institute, Moscow, Russia\\}

\begin{abstract}In these concluding remarks we concentrate on the current state
of variable star studies including halo and globular-cluster
variables, touch on some problems of the distance scale, and propose a
new improvement to the well-known Baade-Wesselink method of determining the
radii of variable stars.
\end{abstract}

\section{Introduction}
B.V.Kukarkin's scientific interests concentrated mainly around
variable stars, globular star clusters, and stellar populations.
His first experience in variable star study as an amateur astronomer
grew with time into a serious engagement into one of the most
impressive Russian and Soviet astronomical projects - the General
catalog of variable stars (GCVS). Kukarkin, together with his colleague and
friend Pavel Parenago, were the initiators of  this long-term
and tedious work of great importance. Nowadays GCVS remains a very popular and valuable knowledge source on different types of variable stars and
accumulates observational data that may serve as a potential source of insight
into some special evolutionary stages of stars and physical processes involved.

In recent years, variable stars have been outlining the frontiers of modern astrophysics
as  "beacons" of stellar evolution. For example,
astro-seismological data are used to test the theories of stellar structure
and evolution; different types of variable stars serve as
indicators of advanced evolutionary stages on the CMD, and
observations of evolutionary period changes in classical Cepheids and secular variability of
post-AGB stars --- predecessors of planetary nebulae ---
provide unique information on stellar evolution. All fields of modern astrophysics demonstrate the need
for variable-star databases.

Serious problems that the authors of the GCVS had to address and
that were associated with the classification of variable stars,
gathering the data and updating the database, and with catalog's
compilation have already been discussed in this meeting by
N.N.Samus. The GCVS now includes more than 60000 entries. Future
difficulties are expected to be more serious than those of the
past. For example, the ASAS-3 project has produced more than 30000
new suspected variables; and GAIA mission will result in
approximately $10^8$ new variables in total -- maybe, it will
detect $\sim~10^4 - 10^5$ new variables every day. Russian space
mission LYRA and many other international all-sky space and
ground-based dedicated deep-sky surveys raise a lot of new
large-scale problems that the GCVS team had not to face until now.
Huge data sets, the ever increasing photometric accuracy
accompanied by the sharp increase in the number of variability
phenomena discovered require fundamentally new solutions, which
must involve close international collaboration. The main features
of the virtual observatory approach (unification of query and
output formats) and the development of new automatic
classification schemes based on self-learning algorithms and
neural networks may prove to be of great importance for solving
future problems.

\medskip

\section{Halo variable stars}
RR Lyrae pulsating variables represent the old stellar population
in galactic halos, thick disks and globular clusters. These stars
populate a narrow region in the CMD - the intersection of cluster
horizontal branch (HB) with the instability strip, and their
energy sources are helium core and hydrogen shell burning. This
stage lasts approximately 100 Myr and can be considered as some
kind of the "second main sequence", by the contrast with the
other, relatively short-lived advanced stages. The luminosities of
RR Lyrae in an individual cluster differ only slightly, with a
characteristic scatter in absolute magnitudes amounting to $\pm
0.15^m$. As was first suggested by Christy (1966) and confirmed by
more recent theoretical track calculations (Vandenbergh et al.
2000), the optical luminosity of HB stars strongly depends on
their chemical abundance, and hence the metallicity $[Fe/H]$ and
$[\alpha/Fe]$ ratio are key parameters. The slope of the $M -
[Fe/H]$ relation was estimated using different methods including
the Baade-Wesselink technique and direct HIPPARCOS and HST FGS3
parallax measurements. Observations appear to support theoretical
predictions and suggest a nearly universal slope for the $M -
[Fe/H]$ relation from optics to NIR (Cacciari and Clementini 2003;
Catelan et al. 2004), with $dM/d[Fe/H] \approx (+0.20 \pm
0.04)^m/dex$. In the NIR  RR Lyrae, like classical and Type II
Cepheids, show a period luminosity ($M - log P$) relation  best
revealed by observations of RR Lyraes in globular clusters (Frolov
and Samus 1998) with a slope of $dM_K/d log P \approx (-2.34 \pm
0.07)^m$.

The statistical-parallax technique offers new opportunities for
the refinement of the luminosities of RR Lyrae variables  (see a
comprehensive review by Gould and Popovski 1998). The above
authors analyzed the eventual biases of this method due to the
kinematical inhomogeneity of the original sample and other
factors. Cacciari and Clementini (2003) noted poor discrimination
of halo and thick disk variables, and a large fraction of
"accretion" population among observed halo stars. Two halo
subsystems would have different dynamical characteristics and
origins: the fast rotating subsystem associated with the Galactic
thick disk, and the slowly (possibly retrograde) rotating
subsystem belonging to the accreted outer halo (Bell et al. 2008).
Any kinematical inhomogeneity in the sample used may introduce
unpredictable systematical errors into the derived distance scale.
The statistical parallax technique, very powerful and robust in
itself, needs more adequate kinematical models for halo
populations and more extensive RR Lyrae samples with good radial
velocities and proper motions, and seems to be a very important
task for future investigations. The kinematics the of local RR
Lyrae population and the associated distance scale was analyzed in
details by Dambis and Rastorguev (2001) and Dambis (2009).
Apparently, the best way to account for the biases of the
statistical parallax technique consists in applying it to
simulated inhomogeneous data sets and estimating the systematical
errors.

Kukarkin used the distances of the globular clusters calculated
from the original Christy's (1966) idea that the luminosity of
stars at the HB stage strongly depends on metallicity (with the
overestimated slope of $dMV/d[Fe/H] = 0.38^m$), to calibrate the
luminosities of the Cepheids in globular clusters (Kukarkin and
Rastorguev 1972, 1973). This relatively poor sample is now
considered to be a mixture of stars of different nature: low-mass
stars evolving from the HB and entering the instability strip
(above horizontal branch variables, AHB), and low-mass stars on
the asymptotic giant branch stage (AGB) looping inside the IS
during the thermal instability phase (true Type II Cepheids). The
above authors noted that the $M_V$ - log P relation has a break
near the 7-8 day period. This is quite similar to what we see in
the case of classical Milky Way and LMC Cepheids, which show a
pronounced slope break near the 10-day period (Sandage et al.
2004). The slope difference may complicate the use of Cepheids as
standard candles. Nonlinear calculations of Cepheid models also
seem to support the idea of two Cepheid families. Unfortunately,
in the last three decades the progress in the detailed studies of
Type II Cepheids and Cepheids in globular clusters was by far not
as impressive as with type-I Cepheids. For example, there is even
a certain confusion regarding the very name of the class of type
II Cepheid variables: some investigators prefer to use the term
short-period Cepheids for BL Her type stars rather than for AHB
Cepheids. We should mention valuable data on Type II Cepheids in
the galactic field, in globular clusters and in nearby galaxies
derived from recent NIR observations (Matsunaga et al. 2006,
2009). These data have substantially expanded the sample Type II
cepheids; the results seem not to confirm early suggestion of the
existence of any PL breaks and demonstrate small scatter of
$JHK_S$ PL relations. The metallicity effect in the luminosities
of Type II Cepheids is now discussed. We expect Type II Cepheids
to be potentially good "standard candle" candidates and useful
tools for estimating distances to halo populations of external
galaxies.

\section{Variable stars as evolution probes}
In many cases stellar variability is a "lighthouse" of stellar
evolution. The instability strip that crossing the entire CMD is the best
argument. Classical Cepheids and other pulsating variables with
relatively stable cycles (RR Lyrae, W Vir, etc.) seem to be good
probes of stellar evolution which is accompanied by the rearrangement
of stellar interiors. It is well known that the lines of constant
periods are not parallel to the evolution tracks, and hence the star MUST
change its pulsation period. The search for secular period changes
was among the first B.V. Kukarkin's scientific activities of the
1930th. Nowadays, Leonid Berdnikov and David Turner made a decisive contribution
to this field  (Turner and Berdnikov 2004; Turner et al. 2006). The new
data obtained as a result of their multicolor photometric monitoring of classical Cepheids
supplemented them by "historical" data "recorded" in old photographic plates allowed
them in many cases to study period changes over 150-years long time intervals.
The above authors reveal secular period changes in many classical
Cepheids, often masked by spurious period variations. The sign and
magnitude of period change rate are unique indicators of the
evolution stage -- the number of the crossing of the instability strip and the
speed of evolution along the track.

Careful inspection of CMD tracks for massive stars shows a number
of successive loops crossing the instability strip with large
differences in Cepheid's luminosity ($\sim 1^m$). It is now widely
recognized that the identification of the crossing number can
reduce appreciably the scatter of the PL relation. The study of period
changes for different types of pulsating variables can be
viewed as a promising way to improve the period-luminosity
relation of Cepheids as the best "standard candles", and universal
distance scale in general. Cepheids and other types of pulsating
variables are not the only objects to exhibit evolutionary
effects; Arkhipova et al. (2007) have recently pointed out that some
supergiants with infrared excess also show very fast evolution
from the AGB to Planetary Nebulae on the CMD, which is accompanied by photometric
and spectroscopic trends over a short time interval (dozen of years).

Great importance of stellar evolution and distance scale studies
makes the scanning and digitizing of historical astro-plates one of
the foremost observational tasks.

\section{Globular clusters and galactic halo populations}
Globular star clusters have always been considered natural
laboratories of stellar evolution and stellar dynamics.
The old idea of globular star clusters as typical examples of simple
stellar population (SSP) has greatly changed in the past years. In the
1960ies, there was no direct observational evidence for the presence of
binary stars among cluster members. Only after 1975, when X-ray emission was
detected from the central parts of densest globular clusters, it occurred
that the explanation may involve binary systems with compact objects.
Now binary population seems to be typical for globular clusters.
Binarity explains the phenomenon of "blue straggler" stars
(BSS). More than 3000 BSS were have been detected in ~60 thoroughly studied
clusters  (Piotto et al. 2004). Peculiarities of the radial distribution
of BSS relative to red-giant stars and the relation between BSS
populations and some of the cluster properties imply two scenarios of their
production: mass exchange in collisional and primordial binaries
(Davies et al. 2004). Large relative frequency of BSS in the halo
field as compared to cluster population is also a very interesting
result. Close binaries seem to be a typical population among
blue horizontal branch (BHB) and extended horizontal branch (EHB)
stars (Ferraro et al. 2001). Note also that the
fraction of binaries in old open clusters is higher than in
globular clusters: according modern data, the binary fraction in old open cluster
cores is estimates at $>11 \%$, whereas the overall fraction ranges from
$35 \%$ to $70 \%$ (Sollima et al. 2009).

Recent photometric and spectroscopic observations with HST and
VLTs revealed multiple main sequences and turnoff-points in some
globular clusters including $\omega Cen, NGC 2808$ etc. (Bedin et al. 2004,
Piotto et al. 2005, Villanova et al. 2007, Piotto et al. 2007).
Even more exciting was the detection of two populations with
different $[Ca/Fe]$ among red giant stars in the nearest "normal"
globular cluster $NGC 6121 = M4$ (Marino et al. 2008). These
observations are indicative of complex star formation processes
in globular clusters, maybe of multiple populations, which can be
closely related to the cluster dynamical history and some special
properties, such as escape velocity.

The old view of globular clusters as a very homogeneous populations in the
Milky Way and external galaxies has been drastically changed by recent
studies of galaxy-formation processes, beginning from very high redshifts. Theoretical
simulations of galaxy formation in the $\Lambda CDM$ scenario (early
clustering of dark matter, accretion of ordinary matter to DM
clumps) have shown that at early epochs multiple mergers -- major
and minor -- took place and eventually shaped the recent "faces" of spiral and
elliptical galaxies.  These calculations also show the formation
of cluster-like clumps (Kravtsov and Gnedin 2005). Recent SDSS
data have clearly shown that minor merger events occur just now,
and stellar traces of the disruption of dwarf galaxies and globular
clusters in the Milky-Way tidal field can be identified among the
thick disk and halo populations (Bell et al. 2008, Koposov and
Belokurov 2008, Smith at al. 2009). The study of peculiarities in the
kinematics and chemical abundances made it possible to identify "accretion"
populations among galactic globular clusters, thick-disk stars and
even among the nearest stars (Marsakov and Borkova 2005ab, Marsakov
and Borkova 2006). Dr. Clementini (see paper in this book)
demonstrated how the populations of RR Lyrae of OoI/OoII types
helps to establish the merger history and understand the Milky-Way formation
via mergers of faint dwarf spheroidal galaxies. The age and abundance
differences between "accreting" clusters originated in dwarf
satellites and "normal" galactic globular clusters could provide an insight
into the so called problem of the "second parameter", which is responsible for the
HB morphology.

The kinematics of distant halo objects -- dwarf Milky-Way
satellites, globular clusters, RR Lyrae variables, constant BHB
stars -- which are easily identifiable among field stars, serve as a tool
to set additional constraints onto the gravitational potential of the Milky Way
and the contribution of dark matter. An important work was done by Battaglia
et al. (2005) who used Jeans equations and radial velocities of
about 250 distant objects to find the best fit for the
dependence of velocity dispersions on Galactocentric radius.
They found the velocity distribution to be dominated by transverse motions at large distances and
estimated the total Milky-Way mass at $\sim 10^{12} M_{sun}$.
Recently Dambis (see paper in this book) used radial-velocity and
proper-motion data (adopted from the last SDSS data release) for a large sample
of halo objects (including BHB stars) to estimate the local Milky Way
rotation velocity and the shape of the velocity ellipsoid out to a
Galactocentric distance of $\sim 40 \; kpc$. He
demonstrated that 3D-velocities are good enough to constrain flat
rotation curve with $V_{rot} \approx 195 \pm 10 \; km/s$ and
confirm large Milky Way mass dominated by DM on large distances.

\section{Distance scale}
The reliability of the universal distance scale is of exclusive
value to all astrophysics, stellar astronomy and cosmology. It is
well known that Cepheids are objects of exceptional importance because
their PL relation makes them good "standard candles" in distant
galaxies. There are a number of ways to determine the
period-luminosity (PL) and period-luminosity-color (PLC) relations
from observational data (see the review of Sandage and Tammann,
2006). It is common practice to scale all distance scales of different "candles"
to the LMC distance. One of the HST key projects was devoted to
precise LMC distance and Hubble constant measurements (Freedman et
al. 2001). As a result, it emerged with "mean" LMC distance
modulus $(m-M)_0 = 18.5 \pm 0.10^m$. Schaefer (2008) analyzed all
LMC distance estimates and concluded that after 2001 this average
value has been generally accepted, and all reported LMC distance
estimates clustered more tightly around the "mean" value, actually too
tightly with an appreciable excess of too precise measurements.
Schaefer used Kolmogorov-Smirnov (K-S) test to demonstrate that this
clustering may be the symptom of a worrisome problem, which is well known
as the "band-wagon" effect or the correlation between papers resulting from
underestimated systematic errors. Schaefer
mentioned that adequate treating of the systematic uncertainties
is one of the serious problems that can greatly affect the
estimates of the distances and other astrophysical parameters.

Given that the calculation of pulsation radii can
improve the luminosity calibrations for Cepheids and RR Lyraes,
I propose a new approximation for the projection factor (PF) used in all
implementations of the W.Baade-W.Becker-A.Wesselink-L.Balona (BBWB)
technique to calculate the radius difference from the radial-velocity
curve. This technique can be used in the surface-brightness method
(Baade 1931) and in the modeling of the light curve -- and also in its
maximum-likelihood implementation (Balona 1977).

The PF value calculated by integrating the flux across the stellar limb
depends on limb darkening coefficient, $\epsilon$ (from
$D(\varphi) = 1 - \epsilon + \epsilon \; cos \varphi, \; \varphi $
is the angle between line-of-sight direction and the normal vector to the surface element),
and the velocity of the photosphere, $dr/dt$. The intrinsic line
profile is broadened by any spectral instrument, and usually we
approximate the profile by a Gaussian curve to measure the radial
velocity as the coordinate of the maximum. This is the standard
technique used by CORAVEL-type spectrographs (Tokovinin 1987). I
used these elementary geometric considerations to show that the
measured radial velocity should additionally depend on the
instrument spectral line width, $S_0$, so PF value should be
"adjusted" to the spectrograph used for radial-velocity
measurements.

We calculated the PF values for $2.5 < |dr/dt| < 60 \; km/s; \; 0 <
\epsilon < 1; \; 4 < S_0 < 8 \;km/s$. We see from Fig.~1 that the
maximum of normal approximation is shifted relative to the tip of
the line profile. This shift depends on $dr/dt$ and $S_0$. Figure~2
shows that the calculated PF values differ considerably from the
"standard" and widely used value of $PF=1.31$. Variations of PF were
reported earlier by Nordetto et al. (2004) and others. The PF
variation with the period, mentioned earlier by some authors,
possibly reflects PF variation as a function of the limb-darkening
coefficient.

\begin{figure}[!h]
\centerline{\hbox{\psfig{figure=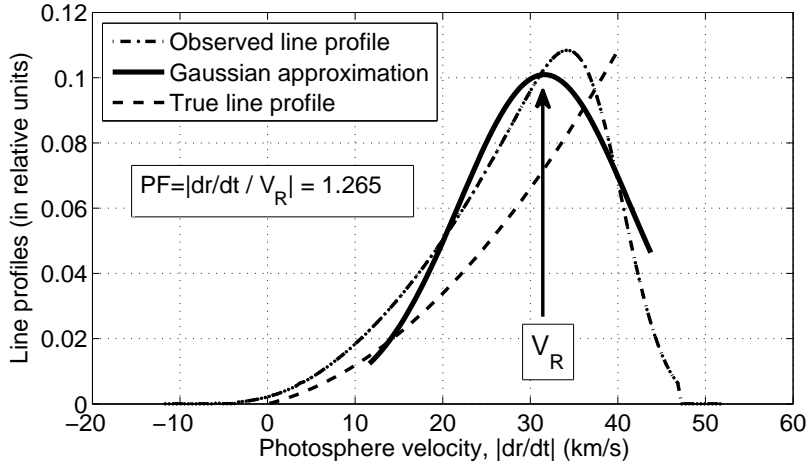,angle=0,clip=,width=12cm}}}
\caption[]{An example: intrinsic line profile ($\epsilon = 0.75,
dr/dt = 40 \; km/s$), observed line profile and its Gaussian
approximation (with the spectrograph instrumental line width $S_0
= 4 \;km/s$). Measured $V_R$ value is indicated by the arrow; the
PF value is shown.} \label{Rastorguev-fig1}
\end{figure}

The bottom panel in Fig. 2 shows the variation of the line width with
the photosphere velocity, and this effect was observed during our
measurements of the correlation profile of bright Cepheids with the ILS
CORAVEL-type spectrometer constructed by Tokovinin (1987).
The upper panel clearly shows large variation of PF values in the
interval from $1.23$ to $1.43$, which does not confirm
Groenewegen's (2007) result that PF can be assumed to be
constant, $PF = 1.26 \pm 0.05$.

\begin{figure}[!h]
\centerline{\hbox{\psfig{figure=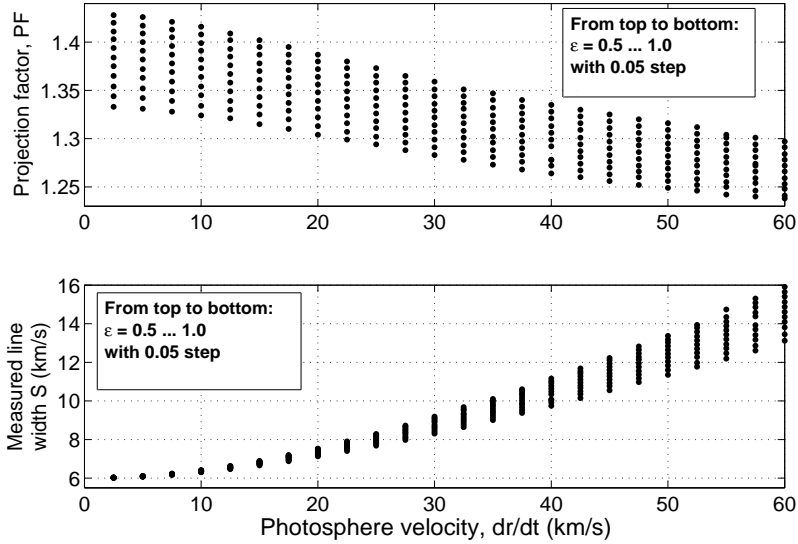,angle=0,clip=,width=12cm}}}
\caption[]{Projection factor, PF, and measured line width, $S$, as
a function of $dr/dt$ for instrumental width $S_0 = 6 \; km/s$ and
for different values of limb darkening coefficient $0.5 < \epsilon
< 1$.} \label{Rastorguev-fig2}
\end{figure}

A useful analytic approximation for the projection factor seems to be
of crucial importance for calculating the radii of Cepheids and RR Lyraes
using the surface brightness technique and Balona's (1977) maximum-likelihood
method of light curve modeling. A careful inspection of PF
as a function of $dr/dt$ leads us to a three-parameter exponential
approximation. Modeling of the line profiles for many sets of
input parameters (see above) allowed us to derive the following
general formula:

\[PF \approx a_1 \cdot exp(-(dr/dt)^2/(2 \cdot a_2^2)) + a_3,\]
where $a_1, a_2, a_3$ are functions of $\epsilon, S_0$:

\(a_1 \approx -0.068 \cdot \epsilon - 0.0078 \cdot S_0 + 0.217\)

\(a_2 \approx +1.69 \cdot \epsilon + 2.477 \cdot S_0 + 9.833\)

\(a_3 \approx -0.121 \cdot \epsilon + 0.009 \cdot S_0 + 1.297\)

The overall RMS residual of the calculated PF value from this analytic
expression is about $\pm0.003$. In the same way, we derived the
following approximation for $S$:

\[S \approx b_1 \cdot (dr/dt)^2 + b_2 \cdot |dr/dt| + b_3,\]
where

\(b_1 \approx -0.001 \cdot \epsilon + 0.00 \cdot S_0 + 0.0027\)

\(b_2 \approx -0.036 \cdot \epsilon - 0.011 \cdot S_0 + 0.152\)

\(b_3 \approx +0.215 \cdot \epsilon + 1.048 \cdot S_0 - 0.743\)

with the RMS of approximation equal to $\pm 0.12$. In practice, the PF
for measured radial velocity should be determined by iterations
for known $S_0$ value.

This work was partly supported by the RFBR grant 08-02-00738.


\begin{thebibliography}{}
\bibitem[Arkhipova et al.(2007)]{Arkhipova et al.2007)}Arkhipova, V.P., Esipov, V.F., Ikonnikova,
N.P. et al. 2007, AstL, 33,604
\bibitem[Baade(1931)]{Baade1931}Baade, W. 1931, Mittel.Hamburg.Sternw., 6, 85
\bibitem[Balona(1977)]{Balona1977}Balona, L. 1977, \mnras, 178, 231
\bibitem[Battaglia et al.(2005)]{Battaglia et al.2005}Battaglia, G., Helmi, A., Morrison, H. et al 2005, \mnras, 364, 433
\bibitem[Bedin et al.(2004)]{Bedin et al.2004}Bedin, L.R., Piotto,
G., Anderson, J et al. 2004, \apj, 605, L125
\bibitem[Bell et al.(2008)]{Bell et al.2008}Bell, E.F, Zucker, D.B., Belokurov, V. et al. 2008, \apj, 680,
295
\bibitem[Cacciari \& Clementini(2003)]{Cacciary&Clementini2003}Cacciari, C., \& Clementini, G. 2003, Stellar Candles for
the Extragalactic Distance Scale (eds. D.Alloin and W.Gieren),
Springer, Lecture Notes in Physics, 635, 105
\bibitem[Catelan et al.(2004)]{Catelan et al.2004}Catelan, M., Pritzl, B.J., \& Smith, H.A. 2004, \apjs, 154, 633
\bibitem[Christy(1966)]{Christy1966}Christy, R.F. 1966, \apj, 144, 108
\bibitem[Dambis \& Rastorguev(2001)]{Dambis&Rastorguev2001}Dambis, A.K., \& Rastorguev A.S. 2001, AstL, 27,108
\bibitem[Dambis(2009)]{Dambis2009}Dambis, A.K. 2009, \mnras, 396, 553
\bibitem[Davies et al.(2004)]{Davies et al.2004}Davies, M.B., Piotto, G., \& de Angeli,
F. 2004, \mnras, 349, 129
\bibitem[Ferraro et al.(2001)]{Ferraro et al.2001}Ferraro, F.R., Diamico, N., \& Possenti, A.
2001, \apj, 561, 345
\bibitem[Freedman et al.(2001)]{Freedman et al.2001}Freedman, V.L., Madore, B.F., Gibson, B.K. et al. 2001, \apj, 553, 47
\bibitem[Frolov \& Samus(1998)]{Frolov&Samus1998}Frolov, M.S., \& Samus, N.N. 1998, AstL, 24, 174
\bibitem[Gould \& Popovski(1998)]{Gould&Popovski1988}Gould, A., \& Popovsky, P. 1998, \apj, 508, 844
\bibitem[Groenewegen(2007)]{Groenewegen2007}Groenewegen, M.A.T.
2007, \aap, 474, 975
\bibitem[Koposov \& Belokurov(2008)]{Koposov&Belokurov2008}Koposov, S., \& Belokurov, V.
2008, Galaxies in the Local Volume (Astrophysics and Space Science
Proceedings), Springer, 195
\bibitem[Kravtsov \& Gnedin(2005)]{Kravtsov&Gnedin2005}Kravtsov,
A.V., \& Gnedin, O.Y. 2005, \apj, 623, 650
\bibitem[Kukarkin \& Rastorguev(1972)]{Kukarkin&Rastorguev1972}Kukarkin, B.V., \& Rastorguev, A.S. 1972, Perem. Zvezdy
Byull., 18, 383
\bibitem[Kukarkin \& Rastorguev(1973)]{Kukarkin&Rastorguev1973}Kukarkin, B.V., \& Rastorguev, A.S. 1973, Variable stars in
globular clusters and in related systems, Proc. IAU Colloq No.21
(eds. J.D.Fernie), Toronto, 1972, 180
\bibitem[Marino et al.(2008)]{Marino et al.2008}Marino, A.F., Villanova, S., Piotto, G. et al.
2008, \aap, 490, 625
\bibitem[Marsakov \& Borkova(2005)]{Marsakov&Borkova2005}Marsakov, V.A., \& Borkova, T.V. 2005,
From Lithium to Uranium: Elemental Tracers of Early Cosmic
Evolution (IAU Symp. Proc., eds Hill, V.; Francois, P.; Primas,
F.), 228, 543
\bibitem[Marsakov \& Borkova(2005)]{Marsakov&Borkova2005}Marsakov, V.A., \& Borkova, T.V.
2005, AstL, 31, 515
\bibitem[Marsakov \& Borkova(2006)]{Marsakov&Borkova2006}Marsakov, V.A., \& Borkova, T.V.
2006, Astron. Astropys. Trans., 25, 149
\bibitem[Matsunaga et al.(2006)]{Matsunaga et al.2006}Matsunaga, N., Fukushi, S., Nakada, Y. et al. 2006, \mnras, 370, 1979
\bibitem[Matsunaga et al.(2009)]{Matsunaga et al.(2009)}Matsunaga, N., Feast, M.W., \& Menzies, J.W. 2009, \mnras, 397, 933
\bibitem[Nardetto et al.(2007)]{Nardetto et al.2007}Nardetto, N., Mourard, D., Mathias, P. et al. 2007, \aap, 471, 661
\bibitem[Piotto et al.(2004)]{Piotto et al.2004} Piotto, G., De Angeli F., King, I.R. et al.
2004, \apj, 604, L109
\bibitem[Piotto et al.(2005)]{Piotto et al.2005}Piotto, G., Villanova, S., Bedin, L.R. et al. 2005,
\apj, 621, 777
\bibitem[Piotto et al.(2007)]{Piotto et al.2007}Piotto, G., Bedin, L.R., Anderson, J. et al.
2007, \apj, 661, L53
\bibitem[Sandage et al.(2004)]{Sandage et al.2004}Sandage, A., Tammann, G.A., \& Reindl, B. 2004, \aap, 424, 43
\bibitem[Sandage \& Tammann(2006)]{Sandage&Tammann2006}Sandage, A., \& Tammann, G.A. 2006, \araa, 44, 93
\bibitem[Schaefer (2008)]{Schaefer2008}Schaefer, B.E. 2008, \aj,
135, 112
\bibitem[Sollima et al.(2009)]{Sollima et al.2009}Sollima, A.,
Carbalo-Bello, J.A., Beccari, F.R. et al. 2009, arXive:0909.1277v1
\bibitem[Smith et al.(2009)]{Smith et al.2009}Smith, M.C., Evans, N.W., Belokurov, V. et al. 2009,
\mnras, 399, 1233
\bibitem[Tokovinin(1987)]{Tokovinin 1987}Tokovinin, A.A. 1987,
SvA, 31, 98
\bibitem[Turner \& Berdnikov(2006)]{Turner&Berdnikov2004}Turner, D.G., \& Berdnikov, L.N.
2004, \aap, 423, 335
\bibitem[Turner et al.(2006)]{Turner et al.2006}Turner, D.G., Abdel-Sabour, A.-L., \& Berdnikov, L.N.
2006, \pasp, 118, 410
\bibitem[Vandenberg et al.(2000)]{Vandenberg et al.2000}Vandenberg, D.A., Swenson, F.J., Rogers, F.J. et al. 2000, \apj, 592, 430
\bibitem[Villanova et al.(2007)]{Villanova et al.2007}Villanova, S., Piotto, G., King, I.R. et al. 2007, \apj, 663, 296

\end{thebibliography}
\end{document}